\documentclass{cernyrep}
\usepackage{texnames}
\usepackage[T1]{fontenc}
\usepackage[bookmarks, colorlinks=true, linktoc=page, pdftex, linkcolor=black, citecolor=black, urlcolor=blue]{hyperref}
\usepackage{hyperref}
\usepackage{fancyhdr}
\fancyhfoffset{4 mm}

\fancypagestyle{ARTTITLE}{%
\fancyhf{} 
\lhead{\small{CERN Accelerator School Proceedings ---
{\it Advanced Accelerator Physics} ---  Spa,  Belgium, 2024}}
\lfoot{\hspace{3mm} Available online at \url{https://cas.web.cern.ch/previous-schools}}
\rfoot{\thepage\hspace*{3mm}}
 
}
\usepackage{varwidth}
\usepackage{subfig}
\usepackage[labelfont=bf]{caption}
\usepackage{xcolor}
\usepackage[sort&compress,numbers]{natbib}

\usepackage{listings}

\begin{document}
\title{Landau Damping}

\author{X. Buffat}

\institute{CERN, Geneva, Switzerland}

\maketitle 

\begin{abstract}
Landau damping is a key mechanism to preserve the stability of particle beams under the influence of various collective forces that would otherwise spoil its quality through beam instabilities. We describe its root cause as well as ways to control it in order to design and operate particle accelerators.
\end{abstract}

\maketitle
\thispagestyle{ARTTITLE}

\section{Introduction}
\begin{figure}
 \begin{center}
  \subfloat[Landau damping~\cite{wikilandau}]{\includegraphics[width=0.4\linewidth]{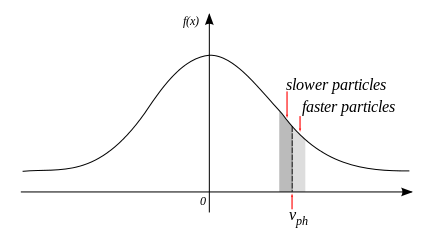}\label{fig-LandauDampingIllustration}}
  \qquad
    \subfloat[Landau anti-damping~\cite{wikitwostream}]{\includegraphics[width=0.4\linewidth]{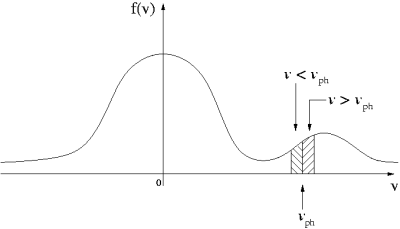}\label{fig-LandauAntiDampingIllustration}}
  \qquad
 \end{center}
\caption{An illustration of Landau damping of a wave with velocity $v_{ph}$ by a distribution of velocities.}\label{fig-LandauAntiDampingIllustration0}
\end{figure}
Landau damping is a somewhat subtle mechanism that occurs when waves supported by an ensemble of particles are damped by an interaction with the motion of the individual particles. To understand the mechanism of Landau damping, it is crucial to realise that the collective motion of an ensemble of particles can significantly differ from the motion of the individual particles that compose it. A good illustration of this difference is given by water waves: They may propagate at velocities in the order of meters per second in a given direction whereas the water molecules that support such waves have velocities in the order of hundreds of meter per second in all direction. The key distinction is that the~macroscopic quantities (here the pressure) are given by statistical properties of the ensemble of particles which do not necessarily reflect the behaviour of the individual particles. In the water wave example, it is clear that velocities of the particles in fact mostly average out, the wave corresponds to a deviation of the average of the velocities. 

In order to illustrate the interaction of a wave with individual particles, we push further the water wave analogy and consider a surfer playing the role of an individual particle. At rest, or even swimming against the wave, the surfer may oscillate in height due to the passage of the wave without gaining or losing energy. However, when the surfer swims in the same direction of the wave he/she is capable of \textit{catching} it and thus increasing his/her speed. By extracting energy from the wave, the surfer contributes to the damping of the wave. Because he/she is alone, his/her impact on the wave is usually not noticeable, but if a fraction of the water molecules that compose the wave were to extract energy in a similar fashion, the wave would be effectively damped. This surfer analogy also highlights the fact that surfer interacts with the collective force, there is no collisions involved. Landau damping is a collisionless process, the~particle-wave interaction occures through collective forces, such as pressure or electromagnetic forces. 

Landau damping occurs in systems where the velocity of the wave and of the individual particles is comparable. Such a configuration is illustrated in Fig.~\ref{fig-LandauDampingIllustration}, where the distribution of the velocities in a~given media is given by a Gaussian and the wave velocity is illustrated with a dashed line. Analogously to the surfer, the particles with a slightly lower velocity than the wave will extract energy from it, whereas particle moving slightly faster will lose their energy to the wave. Since typical velocity distributions (e.g. thermal) feature a lower population at higher velocities, the balance of the energy lost and gained by the~wave is negative thus resulting in damping. Through the same mechanism, more exotic distributions may lead to anti-damping, as illustrated by Fig.~\ref{fig-LandauAntiDampingIllustration}. 

The interaction of a particle with a wave of comparable velocity is analogous the the one of an~oscillator with a wave of comparable frequency. Just like only well-timed pushes to a kid on a swing lead to an increase in oscillation amplitude, the collective force driving the wave may transfer energy to a single oscillator only when the wave has a comparable frequency. Landau damping occurs in a similar fashion as illustrated by Fig.~\ref{fig-LandauAntiDampingIllustration0}: The wave is damped if it is supported by a set of oscillators with a~distribution of frequencies such that there are more oscillators that take energy from the wave rather than give energy to the beam. 

Landau damping was first discovered in plasmas~\cite{landau}, yet the same phenomenon was later found in many other physical systems such as biological clocks (heart beat, fireflies)~\cite{bioclock}, gravitational waves~\cite{landaugravit}, quark-gluon plasma \cite{landauquarkgluon} and of course, particle accelerators. Landau damping is a key component in the~design of particle accelerators, since the interaction of the beam with its surrounding (electromagnetic wakefields, electron-clouds, ions) lead to a self-destructive behaviour called beam instabilities~\cite{instabilities-theseprocs}. While such instabilities can be minimized and sometimes counteracted with active feedbacks, Landau damping is often required to maintain the beam stability. Landau damping is mostly relevant in the design of hadron accelerators, as high enery electron machines usually feature strong synchrotron radiations and thus strong damping of beam oscillations.
\subsection{Decoherence}
\begin{figure}
 \begin{center}
   \subfloat[Turn 0]{\includegraphics[width=0.2\linewidth]{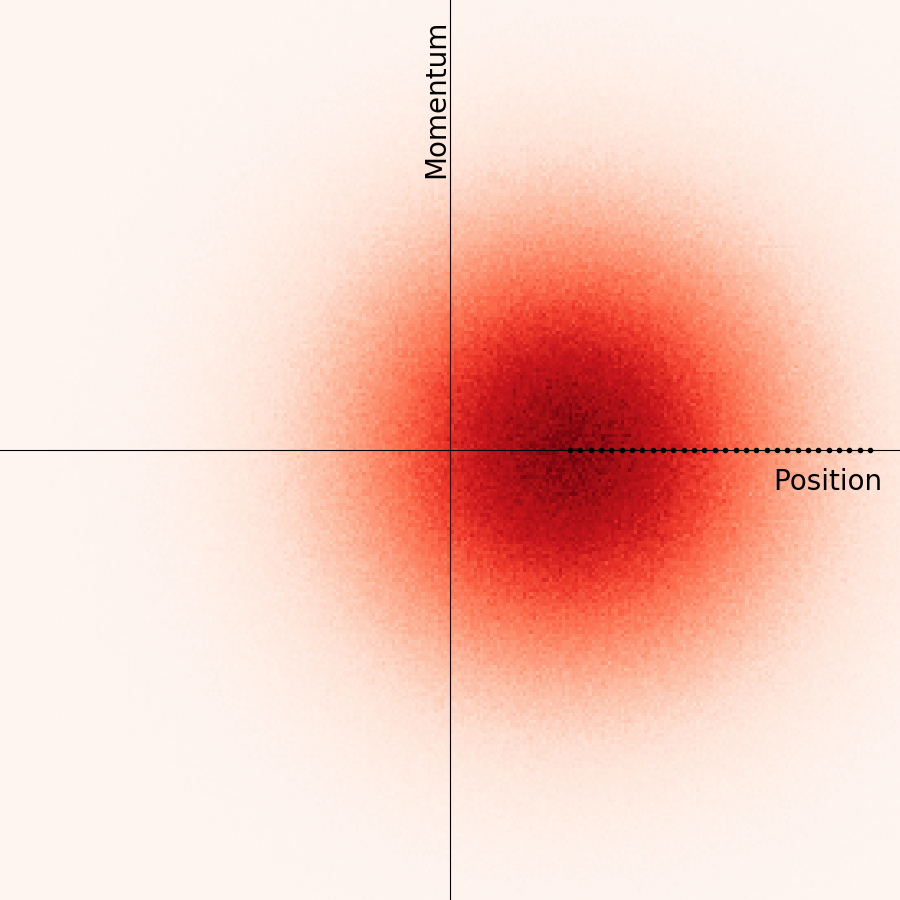}}
  \qquad
  \subfloat[Turn 30]{\includegraphics[width=0.2\linewidth]{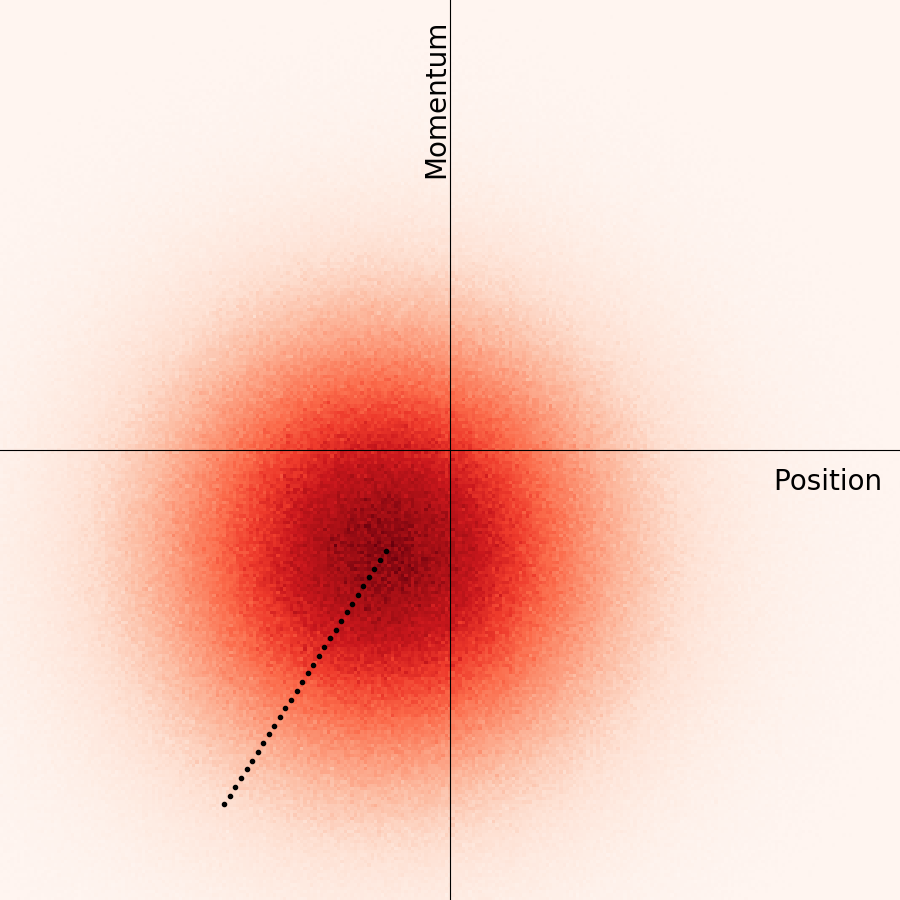}}
  \qquad
    \subfloat[Turn 90]{\includegraphics[width=0.2\linewidth]{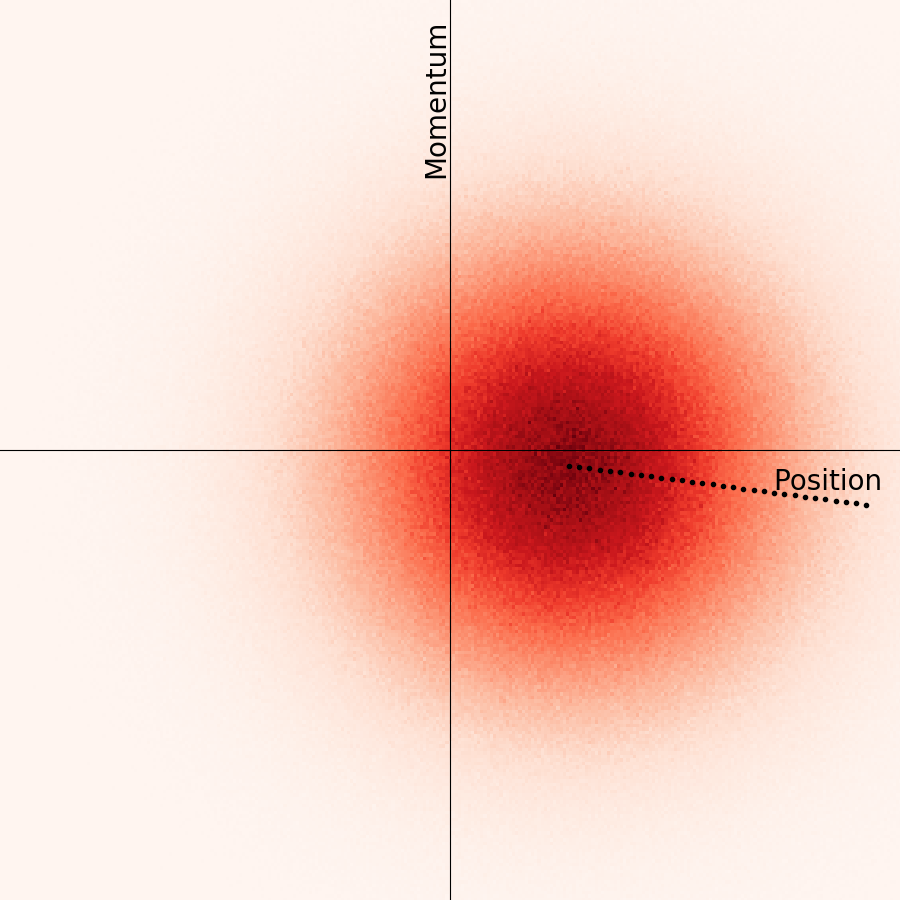}}
  \qquad
    \subfloat[Turn 9000]{\includegraphics[width=0.2\linewidth]{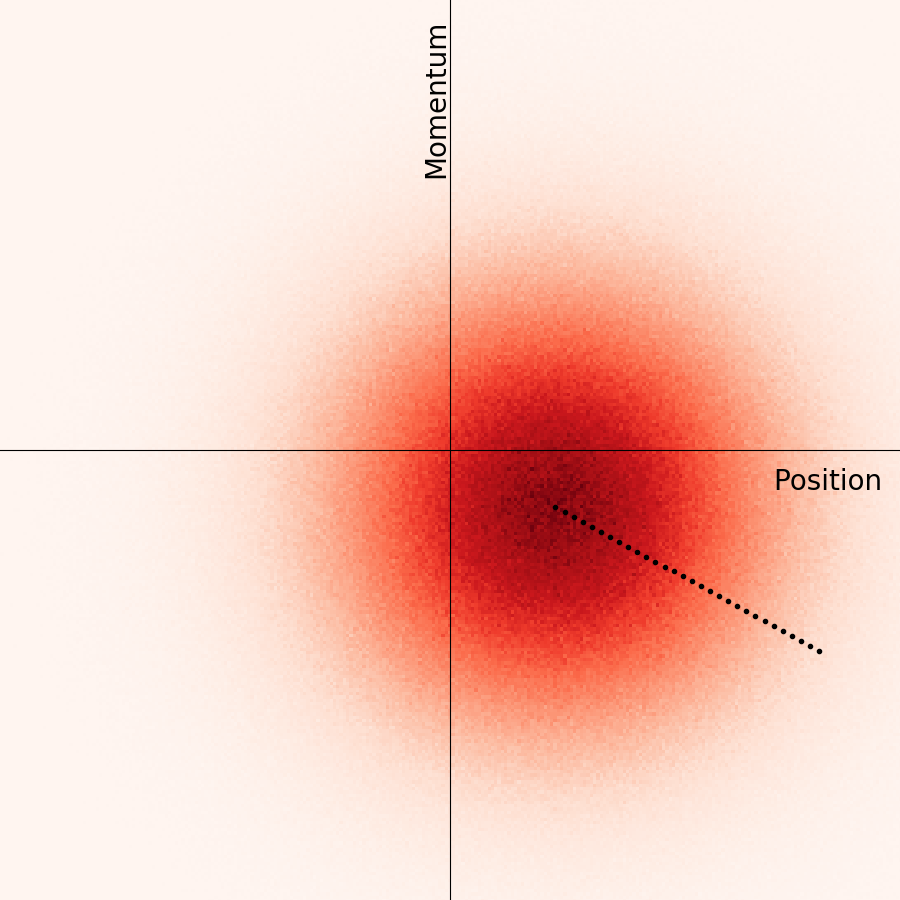}}
  \qquad
  \subfloat[Turn 0]{\includegraphics[width=0.2\linewidth]{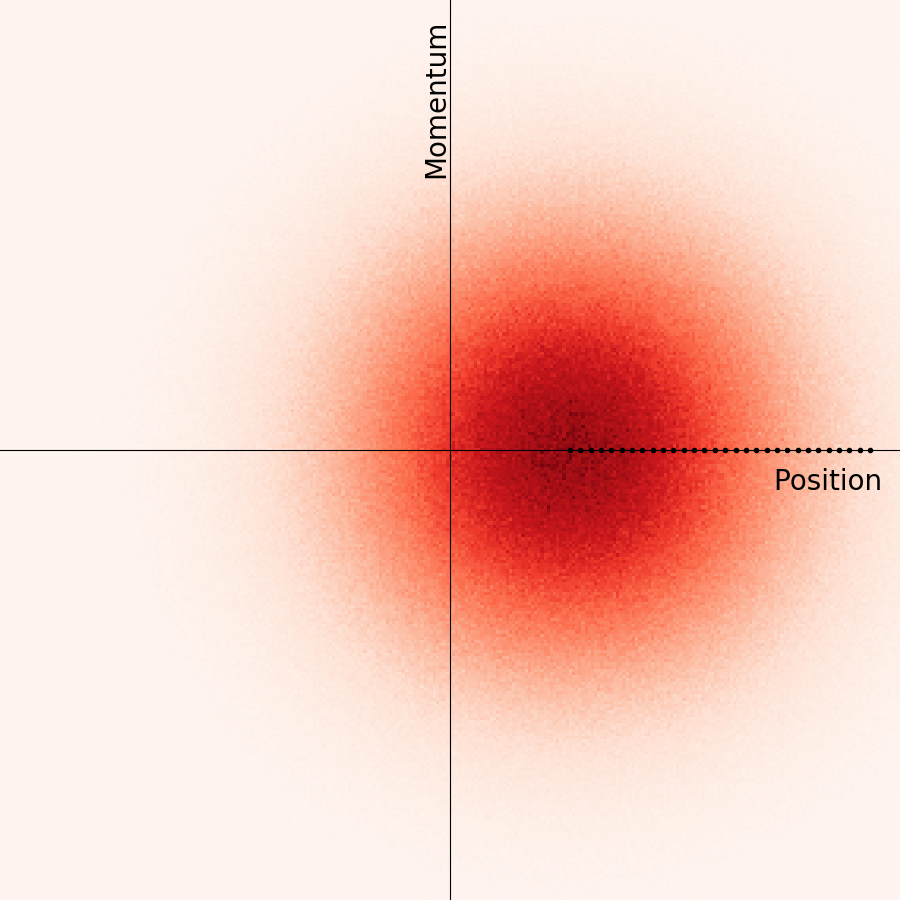}}
  \qquad
  \subfloat[Turn 30]{\includegraphics[width=0.2\linewidth]{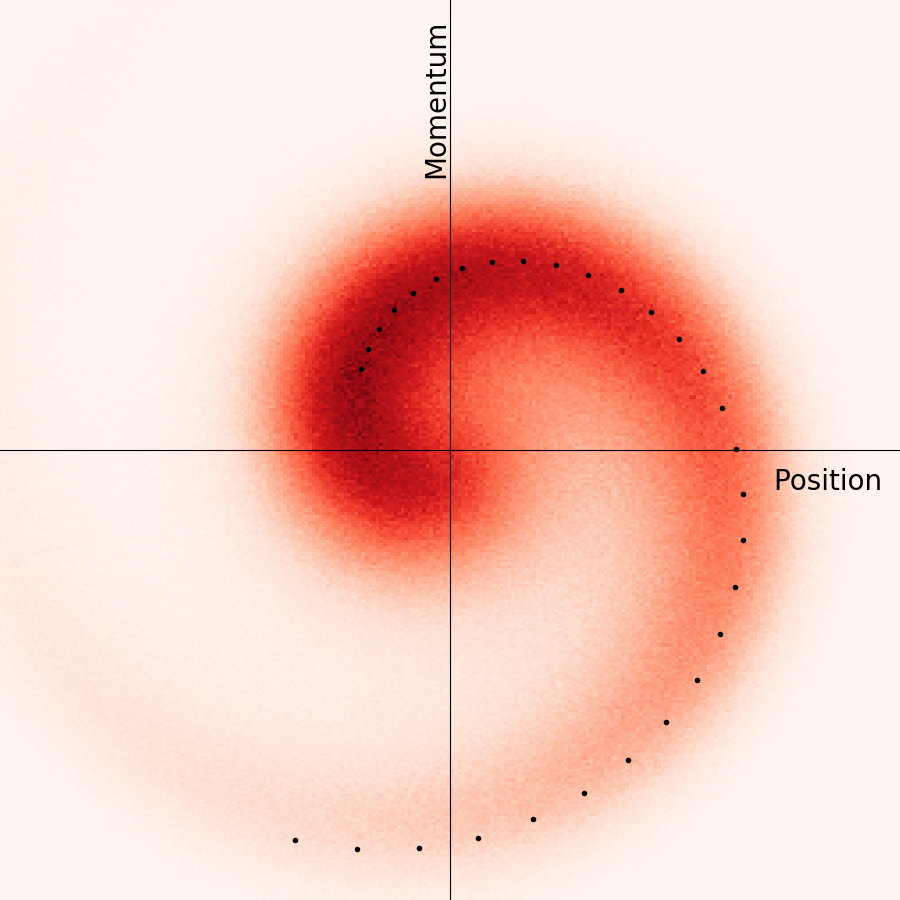}}
  \qquad
    \subfloat[Turn 90]{\includegraphics[width=0.2\linewidth]{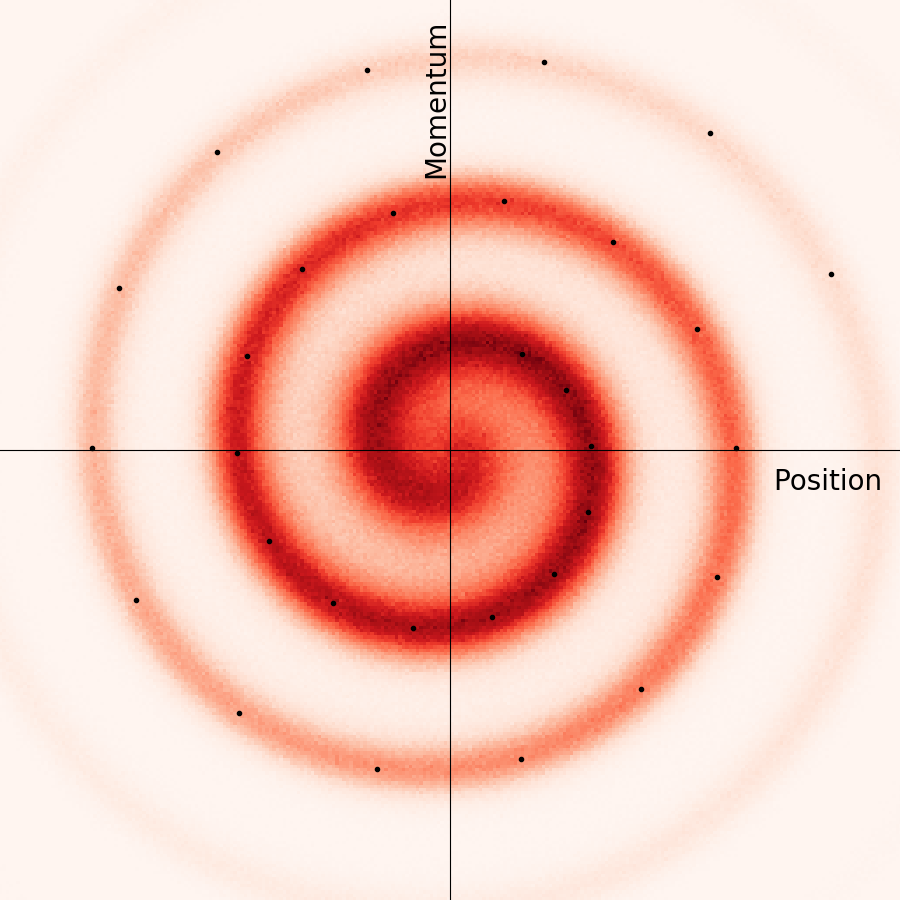}}
  \qquad
    \subfloat[Turn 9000]{\includegraphics[width=0.2\linewidth]{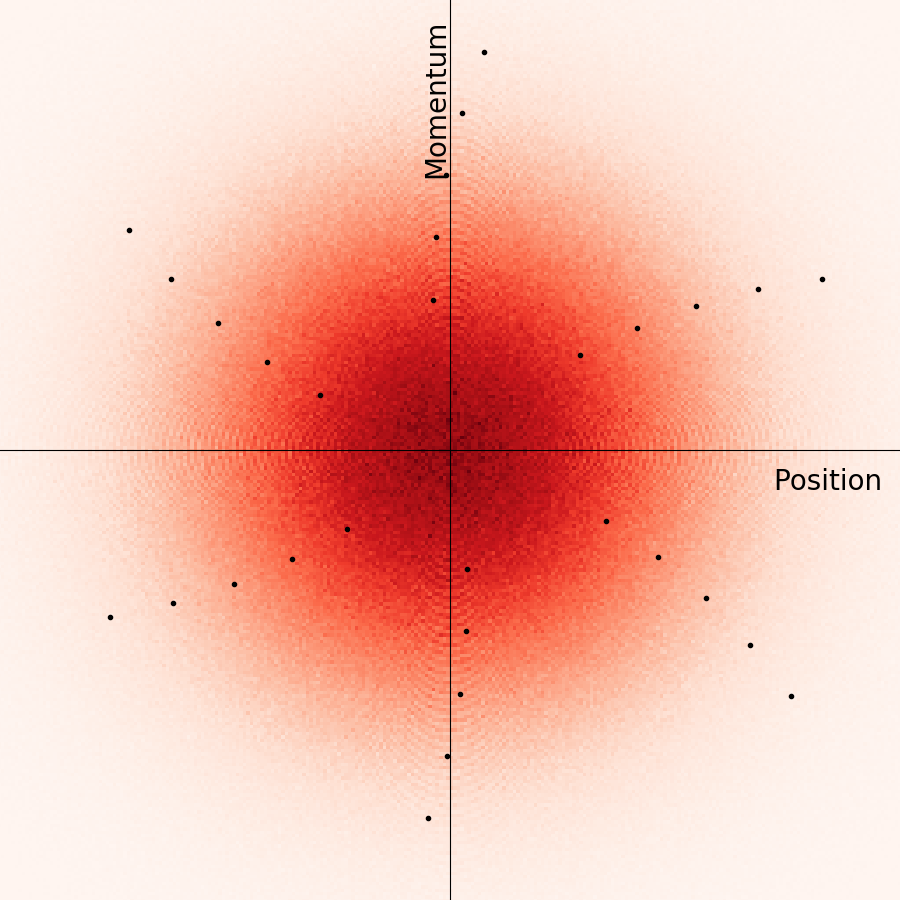}}
  \qquad
 \end{center}
\caption{The mechanism of decoherence illustrated with the evolution of a particle distribution initialised with an~offset position. Few black dots are initialised with different positions to facilitate the visualisation of the~motion. The upper plots feature no detuning, while for the lower plots the oscillation frequency is higher for particles oscillating at higher amplitude.}\label{fig-decoherence}
\end{figure}

In the physics of particle accelerators we usually make a distinction between two closely related phenomenon: Landau damping referes to beams which feature a mechanism of self-amplification that would lead to an exponentially growing oscillation (i.e. a collective instability). The presence of a velocity or a~frequency spread prevents the instability from developping through the mechanism of Landau damping. This occurs without change of the beam distribution, i.e. without emittance growth. We'll focus on this mechanism in the next sections. In the second related phenomenon, an external force (e.g. kickers, ground motion, field ripple, ...) acts punctually on the beam, resulting in an oscillation that is damped at the expense of emittance growth. This phenomenon is usually referred to as decoherence, it is illustrated in Fig.~\ref{fig-decoherence}. A Gaussian distribution of particles is initialised with an offset with respect to the closed orbit. If all particles oscillate with the same frequency (upper plots), the beam oscillate as a whole around the~closed orbit, the collective motion is preserved. However if a frequency spread is introduced, here by increasing the frequency of particles featuring a higher oscillation amplitudes, the particles oscillating at different amplitudes get de-synchronised as time goes. The beam distribution develops a spiralling structure which eventually vanish, leaving a new stationary distribution which is larger than the initial one. In other words, the initial perturbation was damped at the expense of emittance growth. To preserve the~beam emittance, the decoherence is typically minimised by reducing the source of external perturbations acting on the beam. The mechanism of decoherence is further discussed for example in~\cite{CAS-mohl}.
\section{Mathematical formulation}
\begin{figure}
 \begin{center}
  \includegraphics[width=0.6\linewidth]{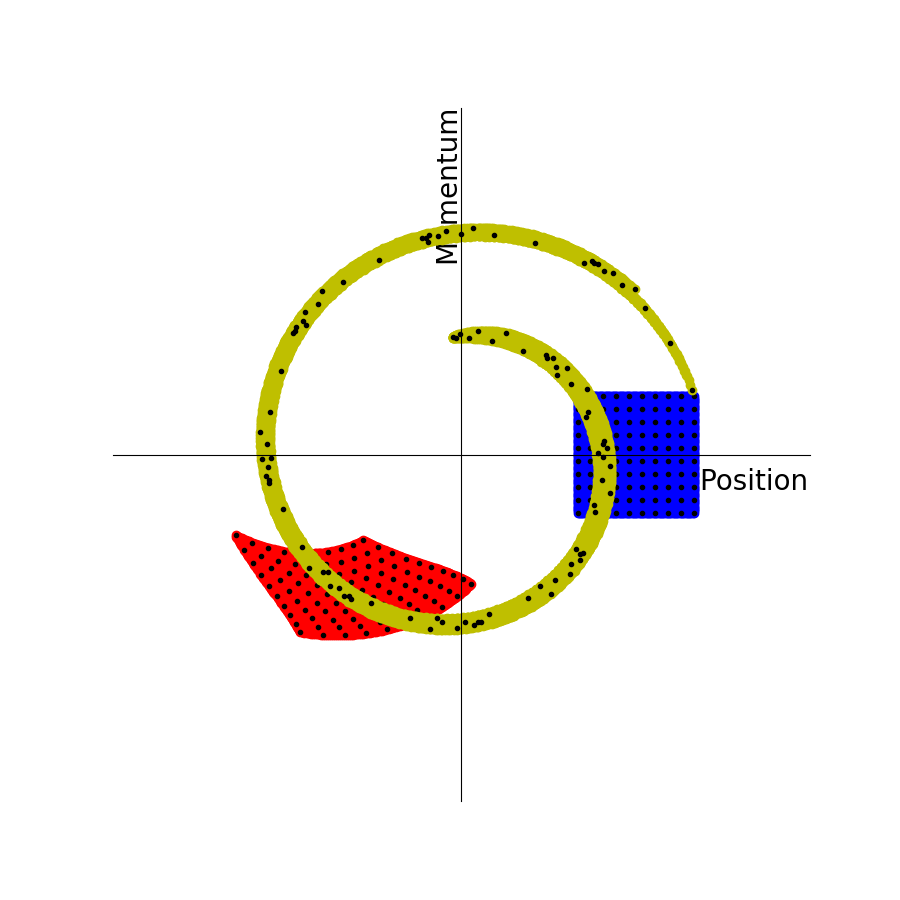}
 \end{center}
\caption{Illustration of the Liouville theorem with a set of particles (black dots) initialised in a blue square. As they evolve in time, the particles later cover the red and finally the yellow area, maintaining its surface. As in Fig.~\ref{fig-decoherence}, the oscillation frequency is higher for particles oscillating with a larger amplitude.}\label{fig-Liouville}
\end{figure}
To illustrate how to quantify the strength of Landau damping in accelerators, we adapt the treatment of Van Kampen~\cite{vanKampen} to particle beams. Equivalently, Landau's approach could also be applied to particle beams~\cite{CAS-herr-landau}. We start from the Liouville theorem~\cite{liouville}, stating the conservation of the phase space density:
\begin{equation}
 \dfrac{d\Psi}{dt} = 0
\end{equation}
with $\Psi$ the particle distribution. The essence of this theorem is illustrated in Fig.~\ref{fig-Liouville}. Considering again the case of a set of particles oscillating around the close orbit with different frequencies depending on their amplitude. In spite of the de-synchronisation of the particles, the phase-space area covered by these particles is preserved, such that the phase space density is constant. Expanding the total derivative using the Hamlitonian $H$, we may write:
\begin{equation}
\dfrac{\partial \Psi}{\partial t} + \sum\limits_i \dfrac{\partial H}{\partial p_i}\dfrac{\partial \Psi}{\partial q_i} - \dfrac{\partial H}{\partial q_i}\dfrac{\partial \Psi}{\partial p_i}  = 0,\label{eq-liouville2}
\end{equation}
where the sum is on the degrees of freedom $i$ with corresponding canonical variables $q_i$ and $p_i$. For simplicity let's consider a single degree of freedom $(x,p_x)$ and a simple Hamiltonian corresponding to the illustrative examples considered above:
\begin{equation}
 H_0 = \omega_0\left(Q_0J+\dfrac{a}{2}J^2\right)\text{, with}~x = \sqrt{2J}\cos(\theta)\text{,}~p_x = -\sqrt{2J}\sin(\theta), \label{eq-NLPerturbation}
\end{equation}
where we have substituted the action-angle variables $J$ and $\theta$ to the Cartesian coordinates $x$ and $p_x$. With this simple Hamiltonian, the oscillation frequency of the particles varies linearly with their action:
\begin{equation}
 \dfrac{\partial H_0}{\partial J} = \omega_0\left(Q_0 +aJ\right) \equiv \omega(J), \label{eq-detuning-oct}
\end{equation}
where we defined the action-dependent frequency $\omega(J)$. Equation~\eqref{eq-liouville2} now reads:
\begin{eqnarray}
\dfrac{\partial \Psi}{\partial t} - \dfrac{\partial H_0}{\partial J}\dfrac{\partial \Psi}{\partial \theta} + \dfrac{\partial H_0}{\partial \theta}\dfrac{\partial \Psi}{\partial J}  &=& 0 \nonumber\\
\dfrac{\partial \Psi}{\partial t} - \omega(J)\dfrac{\partial \Psi}{\partial \theta}  &=& 0.\label{eq-liouville_applied}
\end{eqnarray}
We have used the fact that the Hamiltonian does not depend on the angle to drop the last term. This equation is met for any stationary distribution that does not depend on the angle. This is the case for a~typical beam with a Gaussian distribution
\begin{equation}
 \Psi_0 = \dfrac{1}{2\pi\epsilon_p}e^{-\dfrac{J}{\epsilon_p}}
\end{equation}
with $\epsilon_p$ the beam emittance. Note that the Gaussian distribution in Cartesian coordinates correspond to an exponential distribution in action. In order to study the stability of the system, we consider a first order perturbation of this stationary distribution $\Psi_1(t,J,\theta)$ and a first order perturbation of the Hamiltonian $H_1$, Eq.~\eqref{eq-liouville_applied} equation becomes:
\begin{equation}
 \dfrac{\partial \Psi_1}{\partial t} + \omega(J)\dfrac{\partial \Psi_1}{\partial \theta} - \dfrac{\partial H_1}{\partial \theta}\dfrac{\partial \Psi_0}{\partial J} = 0, \label{eq-liouville_firstOrder}
\end{equation}
where high order terms $\mathcal{O}(H_1^2,\Psi_1^2,H_1\Psi_1)$ were dropped. The perturbation of the Hamiltonian correspond to the collective force that may drive the beam unstable. It may take many forms following the complexity of the phenomenon considered, such as the electromagnetic wakefields, electron clouds, beam-beam, space-charge, ions, ... Based on a given expression of the force $F_{ext}$, we find an expression for the derivative of the corresponding Hamiltonian with $\theta$:
\begin{eqnarray}
  \dfrac{\partial H_1}{\partial x} &=& -F_{ext} \\
  \dfrac{\partial H_1}{\partial \theta} &=& -\dfrac{\partial x}{\partial \theta}F_{ext} \\
  \dfrac{\partial H_1}{\partial \theta} &=& -\sqrt{2J}\sin(\theta)F_{ext}
\end{eqnarray}
such that we may write Eq.~\eqref{eq-liouville_firstOrder} as:
\begin{equation}
 \dfrac{\partial \Psi_1}{\partial t} + \omega(J)\dfrac{\partial \Psi_1}{\partial \theta} - \sqrt{2J}\sin(\theta)F_{ext}\dfrac{\partial \Psi_0}{\partial J} = 0. \label{eq-vlasov_firstOrder}
\end{equation}
This equation is often called the first order Vlasov equation, referring to its analogue in plasmas~\cite{vlasov}. To illustrate the phenomenon of Landau damping we consider a simple collective force proportional to the~average position of the particles in the beam:
\begin{equation}
 F_{ext} = -2\Delta\Omega_{ext}\langle x \rangle,
\end{equation}
it is characterised by the complex frequency shift that it induces $\Delta\Omega_{ext}$. We look for harmonic solutions to the first order Vlasov equation in the form
\begin{equation}
 \Psi_1 = g(J)e^{i(\theta-\Omega t)},
\end{equation}
Equation~\eqref{eq-vlasov_firstOrder} becomes (after a few manipulations)
\begin{equation}
 (\Omega-\omega)g = \dfrac{-1}{2}\Delta\Omega_{ext}\dfrac{df_0}{dJ}\sqrt{2J} \int dJ \sqrt{2J}g. \label{eq-lin-pert-vlasov}
\end{equation}
While Landau studied the solutions to this integral equation using Laplace transforms, we rather follow Van Kampen who later found solutions in terms of distributions rather than functions. The approach is in all aspects standard for integral equations except that distributions are more general mathematical objects with respect to functions: they may feature singularities that cannot be represented by functions, yet their integrals need to remain well defined. This aspect is key, since eventually we are only interested in integrals over the distributions, for example the average position of the beam
\begin{equation}
\langle x \rangle = \int dJ d\theta x\Psi_1. 
\end{equation}
We observe that any solution of Eq.~\eqref{eq-lin-pert-vlasov} can be scaled arbitrarily, we may therefore chose a solution that satisfies
\begin{equation}
\int dJ \sqrt{2J}g = 1. \label{eq-vlasov-scaling}
\end{equation}
such that the first order Vlasov equation greatly simplifies and we may find two sets of solutions. The~first one, usually called the coherent mode, reads
\begin{equation}
 g_c = \dfrac{-1}{2}\Delta\Omega_{ext}\dfrac{\sqrt{2J}\dfrac{df_0}{dJ}}{\Omega_c-\omega}.
\end{equation}
The frequency of the collective mode $\Omega_c$ is found based on the condition Eq.~\eqref{eq-vlasov-scaling}:
\begin{equation}
 \int dJ\dfrac{J\dfrac{df_0}{dJ}}{\Omega_c-\omega} = \dfrac{-1}{\Delta \Omega_{ext}}. \label{eq-dispersion}
\end{equation}
This equation is called the dispersion relation, as it links the frequency shift caused by the collective force (e.g. wakefields) $\Delta\Omega_{ext}$ to the frequency of the coherent mode $\Omega_c$. When considering plane waves in plasmas, Van Kampen showed that all other solutions to the first order Vlasov equation can be expressed as so-called \textit{Van Kampen modes}. The contribution of these modes to macroscopic observables eventually vanish, such that the long term behaviour of the full dynamical system is well described by the one of the coherent modes. Assuming that this also holds for our linearised Vlasov equation, the stability of the~beam is therefore given by the stability of the coherent solution
\begin{equation}
 \Psi_1 = g_c(J)e^{i(\theta-\Omega_c t)}.
\end{equation}
If the frequency of the coherent mode has negative imaginary part, the solution is unstable. Consequently, in order to know whether the beam is unstable, it is sufficient to solve the dispersion relation (Eq.~\eqref{eq-dispersion}) for $\Omega_c$.
\subsection{Stability diagram}
 \begin{figure}
 \begin{center}
  \includegraphics[width=0.6\linewidth]{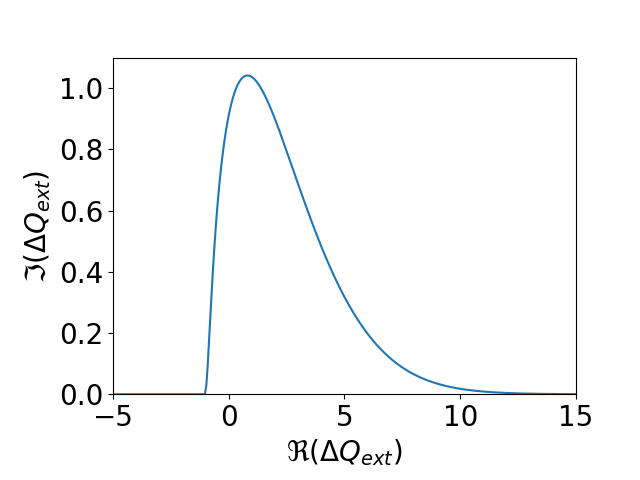}
 \end{center}
\caption{Stability diagram for a linear detuning in one dimension (Eq.~\eqref{eq-SD-1D}).}\label{fig-stabilityDiagram0}
\end{figure}
The dispersion relation is not directly solvable, it is convenient to quantify the impact of Landau damping by drawing the stability boundary in terms of the strength of the collective force. In other words, we find the set of complex frequency shifts $\Delta Q_{ext}$ which give a solution featuring a vanishing imaginary part
\begin{equation}
 \Omega_c = \omega - i \epsilon,~\omega\in\mathbb{R}.
\end{equation}
Considering the configuration discussed above, namely a linear detuning with the action and a Gaussian distribution of particles the dispersion relation can be writen as~\cite{landau2dtunespread}:
\begin{equation}
 \dfrac{-1}{\Delta Q_{ext}} = 1-qe^qE_1(q),~q \equiv \dfrac{Q_c-Q_0}{a\epsilon_p} \label{eq-SD-1D}
\end{equation}
with $E_1$ the exponential integral. We can draw the so-called stabiltiy diagram representing the area in the complex plane of acceptable collective forces. As the line represents coherent modes with vanishing complex frequencies, we infer that stronger collective forces (above the curve) will make the beam unstable, weaker collective forces (below the curve) will maintain the beam stability. We note that configurations featuring a complex tune shift $\Delta Q_{ext}$ below the curve does not necessarily mean that the~complex part of the coherent frequency $\Omega_c$ is negative. Indeed in some areas of the~complex plane the~dispersion relation (Eq.~\eqref{eq-dispersion}) does not feature any solution. As a result, the coherent mode does not exist and the motion is described only by Van Kampen modes. In terms of behaviour, this corresponds also to a stable beam. \\
In the next section, we will study the stability diagrams for few accelerator applications.

\begin{figure}
 \begin{center}
  \subfloat[Beam transfer function]{\includegraphics[width=0.4\linewidth]{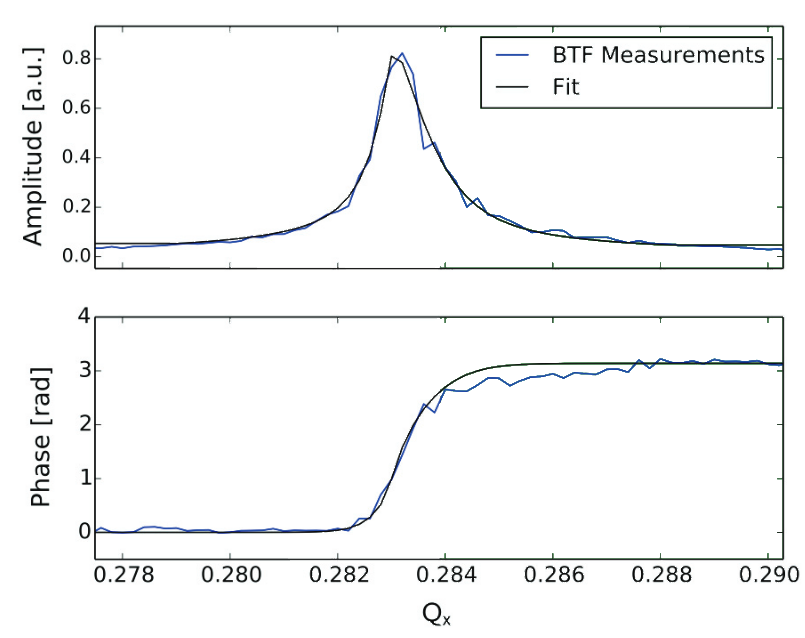}\label{fig-BTFLHC}}
  \qquad
    \subfloat[Stability diagram]{\includegraphics[width=0.4\linewidth]{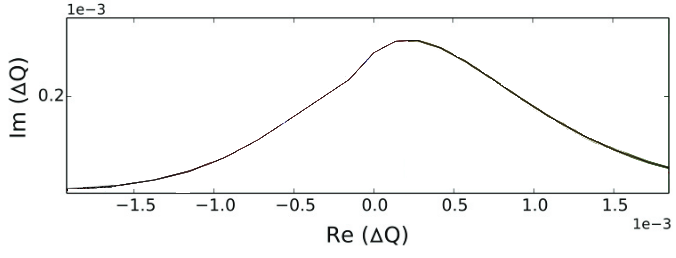}\label{fig-SDfromBTF}}
  \qquad
 \end{center}
\caption{Beam transfer function measured at the LHC and its corresponding stability diagram~\cite{tambasco}.}\label{fig-BTF-SD}
\end{figure}

While the stability diagram may appear as an abstract quantity, it can be measured through the~so-called Beam Transfer Function (BTF). To show this analogy, we go throught similar derivations, yet neglecting collective forces and rather introducing an external harmonic excitation
\begin{equation}
 F_{ext} = A_{d}e^{-i\Omega_d t}
\end{equation}
we can start again from Eq.~\eqref{eq-vlasov_firstOrder}:
\begin{equation}
 \dfrac{\partial \Psi_1}{\partial t} + \omega(J)\dfrac{\partial \Psi_1}{\partial \theta} - \sqrt{2J}\sin(\theta)A_{d}e^{-i\Omega_d t}\dfrac{\partial \Psi_0}{\partial J} = 0
\end{equation}
to obtain
\begin{equation}
  \dfrac{g_d}{A_{ext}} = \dfrac{1}{2}\dfrac{\sqrt{2J}\dfrac{df_0}{dJ}}{\Omega_d-\omega(J)} \label{eq-BTFtoSD}
\end{equation}
and finally
\begin{equation}
 \dfrac{\left\langle x\right\rangle}{A_{ext}}= \int dJ\dfrac{J\dfrac{df_0}{dJ}}{\Omega_d-\omega(J)}.
\end{equation}
We recognize Eq.~\eqref{eq-dispersion} where the collective force, which is usually poorly known and difficult to control, is replaced by a driving force that can be finely tuned. Such a measurement of the amplitude and phase response of the LHC beam to a sinosoidal excitation is shown in Fig.~\ref{fig-BTFLHC}. The response is very much comparable to the~one of a driven damped harmonic oscillator, the damping force at stake is Landau damping. Using Eqs.~\eqref{eq-BTFtoSD} and \eqref{eq-dispersion} we can reconstruct the stability diagram (Fig.~\ref{fig-SDfromBTF}). It defers from the~one shown in Fig.~\ref{fig-stabilityDiagram0} which was derived for a one dimensional tune spread, whereas in the LHC the~bi-dimensional (x and y) nature of the detuning can not be neglected. The derivation of the stability diagram with the two degrees of freedom can be found in \cite{landau2dtunespread}.
\section{Selected examples of application}
\subsection{Unbunched beams}
\subsubsection{Beam intensity modulation}
\begin{figure}
 \begin{center}
 \includegraphics[width=0.6\linewidth]{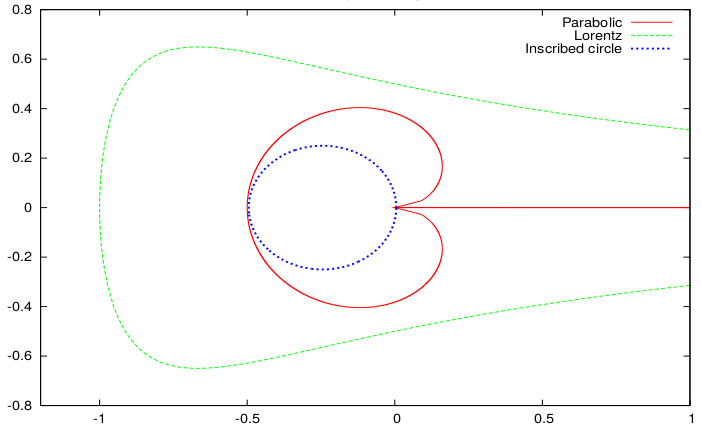}
 \end{center}
\caption{Stability diagram for coasting beams with different velocity distributions and the conservative assumption for the Keil-Schnell criterion (dashed blue circle) \cite{CAS-herr-landau}.}\label{fig-keilschell}
\end{figure}

In a coasting beam, a modulation of the beam density may self-enhance through electromagnetic wake fields. With a model of the longitudinal beam coupling impedance $Z^{\parallel}(\omega)$, on may obtain the frequency shift caused by the electromagnetic wakefields using perturbation theory~\cite{chaoBook}:
\begin{equation}
\Delta\Omega_n = i\dfrac{2\pi N r_0 n\eta}{\gamma T_0^3}Z^{\parallel}(n\omega_0) \label{eq-freqshift_long_coasting}
\end{equation}
with $n$ the number of period of the intensity modulation, $N$ the number of particles in the beam, $r_0$ the classical radius of the beam particles, $\eta$ the slippage factor, $\gamma$ the relativistic factor, $\omega_0$ the revolution frequency and $T_0$ the revolution period. This instability can be stabilised through Landau damping thanks to the spread in revolution frequencies of the particles with different momentum in the beam. In this case the dispersion takes a special form~\cite{chaoBook}:
\begin{equation}
 \dfrac{1}{\Delta \Omega_{n}} = \int d\omega\dfrac{\rho(\omega)}{\left(n\omega-\Omega\right)^2}
\end{equation}
with $\rho(\omega)$ the distribution of revolution frequencies. The corresponding stability diagram is shown in Fig.~\ref{fig-keilschell} for a parabolic and a Lorenzian distribution of revolution frequencies. The impact of the distribution is significant whereas is it often poorly known. It is convenient to consider the Keil-Schnell criterion, which pessimistically considers a circle inscribed into the dervied stability diagram yielding~\cite{keilschnell}
\begin{equation}
|\Delta \Omega_{n}| \lessapprox \dfrac{1}{4}n^2\Delta\omega^2.
\end{equation}
This provides an important criterion to design an accelerator as it constrains for example the machine impedance, the beam intensity and the momentum spread, as $\Delta \omega \approx \omega_0|\eta|\Delta\delta$.
\subsubsection{Microwave instability}
In bunched beams, when the wavelength of the intensity modulation is much shorter than the bunch length, we may neglect the impact of the longitudinal focusing and apply the same stability criterion. The average beam current should however be replaced by the peak current. It is usually referred to as the~Keil-Schnell-Boussard criterion~\cite{boussard}.
\subsubsection{Transverse instability}
In a similar fashion, a modulation of the transverse position around the ring may grow exponentially when interacting with the machine transverse impedance $Z^{\perp}(\omega)$. From perturbation theory one obtains the complex tune shift for those modes of oscillation~\cite{chaoBook}:
\begin{equation}
 \Delta\Omega_n = -i\dfrac{N r_0 c^2\eta}{2\gamma\omega_{\beta} T_0}Z^{\perp}(n\omega_0+\omega_{\beta}).
\end{equation}
The dispersion integral now takes the form~\cite{chaoBook}
\begin{equation}
 \dfrac{-1}{\Delta \Omega_{n}} = \int d\omega\dfrac{\rho(\omega)}{\omega-n\omega_0-\Omega}
\end{equation}
with $\rho(\omega)$ the spread in transverse oscillation frequencies. The transverse distribution of frequency depends on the revolution frequency but also on the chromaticity of the machine denoted $Q'$. As for the~longitudinal case, this distribution is usually poorly known and we rather rely on a conservative simplified stability diagram, such that eventually we obtain the stability criterion
\begin{equation}
 |\Delta \Omega_{n}| \lessapprox \Delta \omega \label{eq-criterion-trans-coast}
\end{equation}
with the frequency spread given by $\Delta \omega = \omega_0|Q'-n\eta|\Delta\delta$.
\subsection{Bunched beams}
In bunched beams the revolution frequency usually plays a much less important role, as particles are forced to revolve around a fixed frequency by the RF cavities. While one may often still apply the simplified criterion Eq.~\eqref{eq-criterion-trans-coast} for the so-called weak head-tail instability~\cite{laclare}, the frequency spread now originate mostly from the non-linearity of the forces that act on the particles. As a result of the non-linearity, particles oscillating with different amplitudes will effectively be focused with a different strength thus leading to different oscillation frequencies. The distribution of oscillation amplitude, or the distribution of actions, then leads to a distribution of frequencies. Let us consider two examples.
\subsubsection{Non-linear RF fields and Landau cavities}
\begin{figure}
 \begin{center}
 \subfloat[Small bucket area]{
 \includegraphics[width=0.4\linewidth]{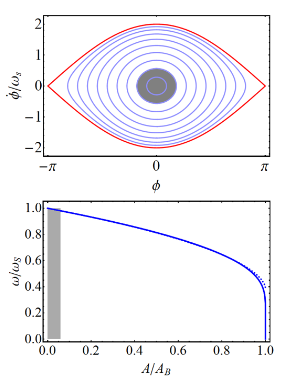}}
 \qquad
  \subfloat[Small bucket area]{
 \includegraphics[width=0.4\linewidth]{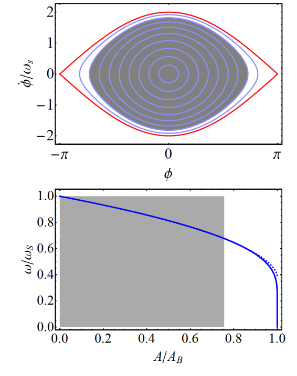}}
 \end{center}
\caption{Phase space trajectories of particle with different oscillation amplitude in the longitudinal phase space (upper plots) with the corresponding frequency variation (bottom plots). The grey shaded areas represent a beam with a small emittance with respect to the available bucket (left plots) and a beam filling well the bucket (right plots) \cite{CAS-heiko}.}\label{fig-longitudinal-phase-space}
\end{figure}
The longitudinal focusing is usually realised with RF cavities providing sinusoidal field. While the sine wave is close to a linear restoring force at the center, the force vanishes for particles that are either very early or very late with respect to the synchronous particle. The resulting trajectories are illustrated in Fig.~\ref{fig-longitudinal-phase-space} (top plots). The trajectories close to the center are about circular, corresponding to a linear motion. As the amplitude of oscillation of a particles increase, the trajectories become elongated, until a separatrix is reached (red curve) beyond which the particles are no longer oscillating around the fixed point, leading to so-called uncaptured beam. The corresponding oscillation frequency is shown in the lower plots. As the amplitude of oscillation increases, the particles probe the decaying part of the sine wave and therefore experience less focusing. As a result, their frequency of oscillation decreases towards 0 at the separatrix. The grey areas represent the trajectories of the particles in a beam with a low longitudinal emittance (left) and large longitudinal emittance (right). In order to profit at most of the Landau damping offered by this mechanism, one must carefully design the RF system such that the targeted beam properties correspond to the situation on the right. Indeed as seen on the bottom plot, the particles will have a larger frequency spread and thus offer stronger Landau damping.

Several machines are equipped with an active excitation mechanism that allows to artificially increase the longitudinal emittance and thus maintain Landau damping in critical phases of the cycle. This reduction of the beam quality is often preferable to beam instabilities that would otherwise deteriorate the beam quality in an uncontrolled way.

\begin{figure}
 \begin{center}
 \subfloat[Bunch lengthening mode]{
 \includegraphics[width=0.4\linewidth]{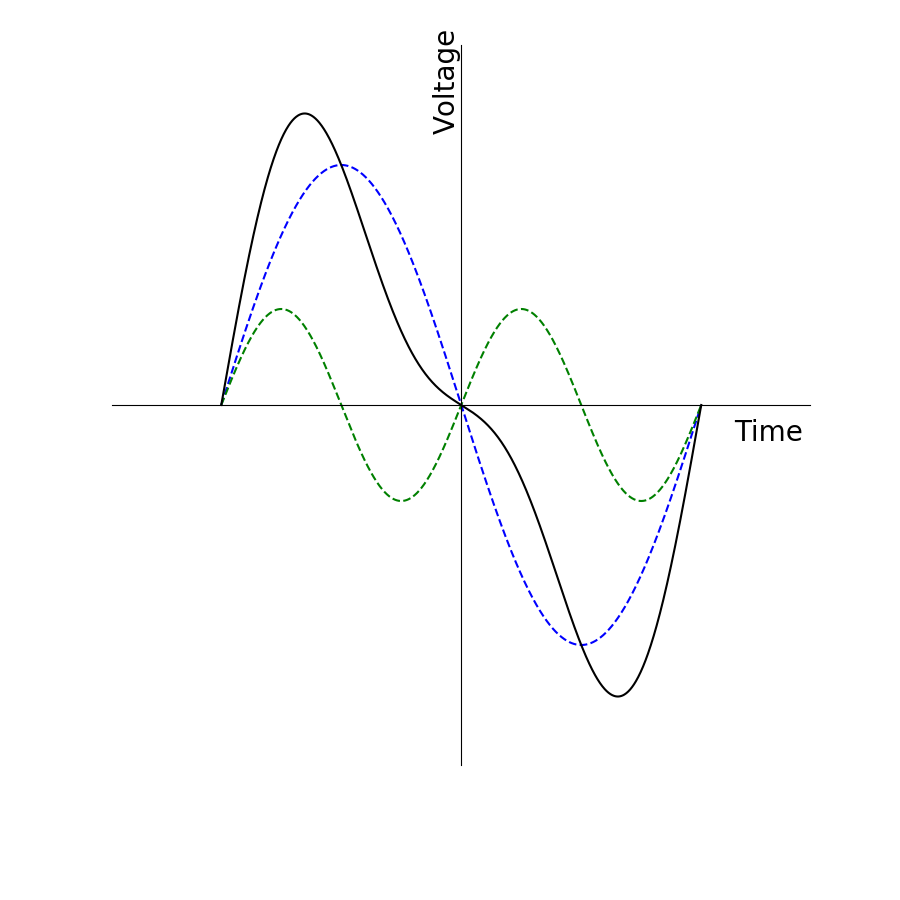}\label{fig-lengthening}}
 \qquad
  \subfloat[Bunch shortening mode]{
 \includegraphics[width=0.4\linewidth]{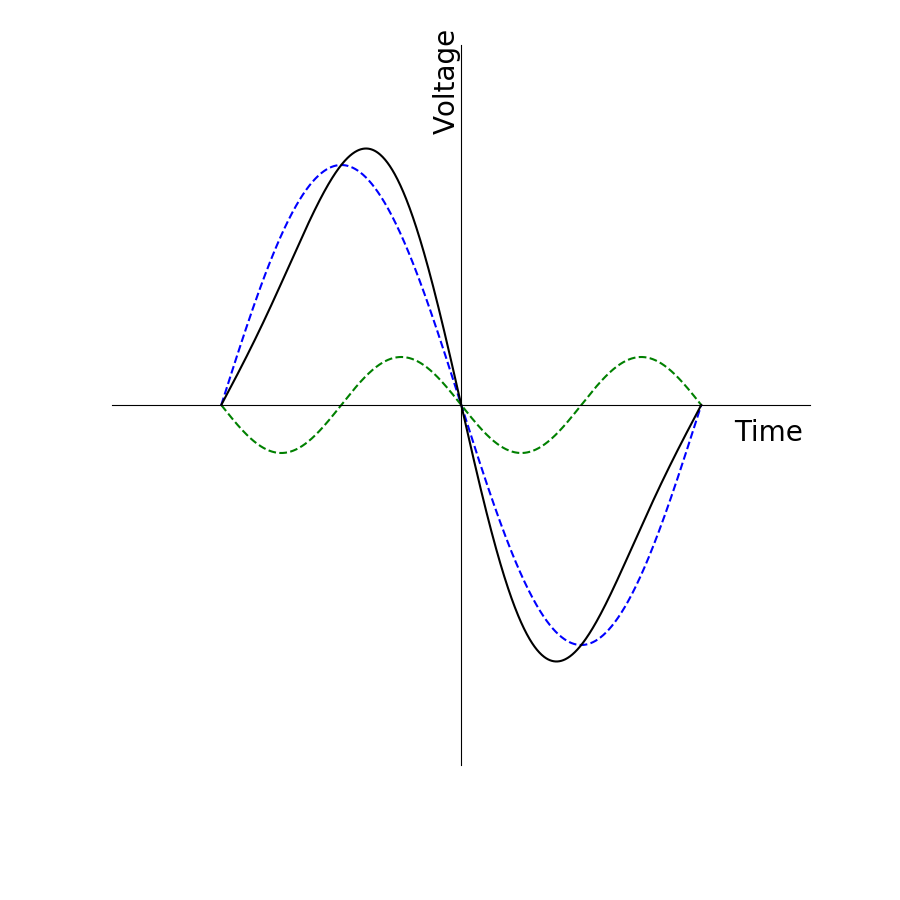}\label{fig-shortening}}
 \end{center}
\caption{Addition of a main RF field (blue dashed line) with a harmonic cavity (green dashed line) forming a~non-linear total voltage (black solid line).}\label{fig-double RF}
\end{figure}

When this natural source of Landau damping is not sufficient to reach the desired beam parameters, it is possible to further enhance it using additional cavities, so-called \textit{Landau cavities}. The purpose of the~cavities is to add non-linear contributions to the main RF field such as to increase the frequency spread and thus Landau damping. A typical setup is the usage of a cavity at a higher harmonic ($\omega_{Landau} = n\omega_0$) with a fraction of the main voltage. If the Landau cavities are out of phase with the main cavities (Fig.~\ref{fig-lengthening}), the focusing is weakened at the center and strengthened at high amplitude. On the contrary when the~Landau cavities are powered in phase (Fig.~\ref{fig-shortening}), the focusing is strengthened and varies more strongly towards the edge of the separatrix. Both may be used with pros and cons~\cite{CAS-heiko}. 
\subsubsection{Transverse tune spread and Landau octupoles}
\begin{figure}
 \begin{center}
 \subfloat[Quadrupole field]{
 \includegraphics[width=0.4\linewidth]{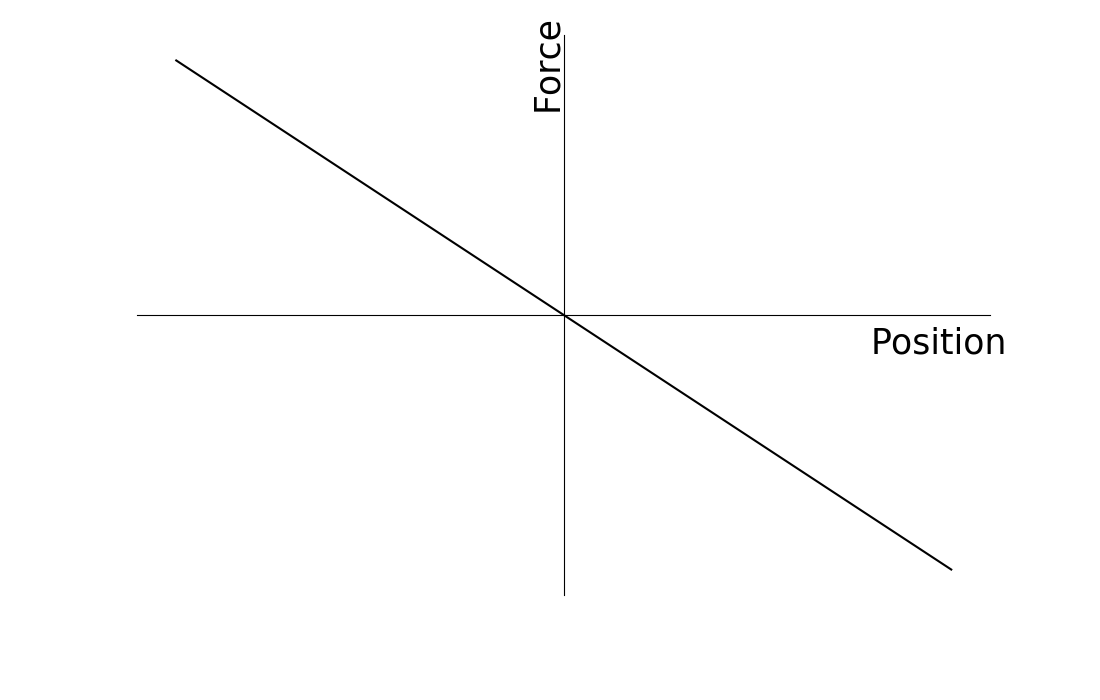}\label{fig-quad}}
 \qquad
  \subfloat[Octupole field]{
 \includegraphics[width=0.4\linewidth]{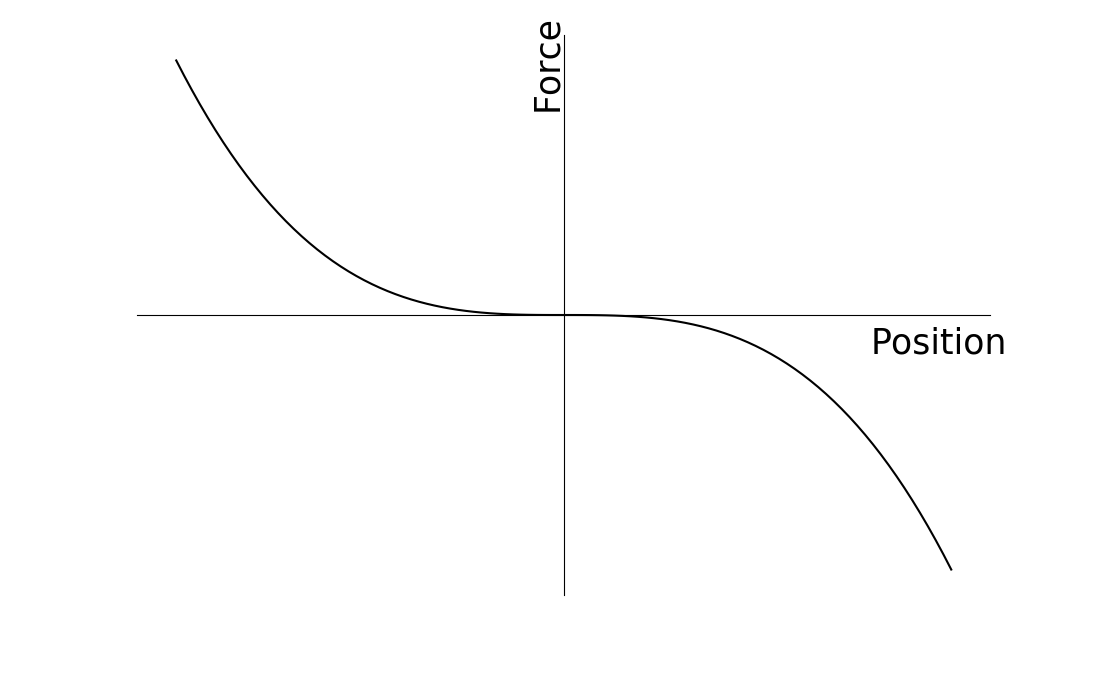}\label{fig-oct}}
 \end{center}
\caption{Comparison of the linear restoring force of a quadrupole magnet and the non-linear force of an~octupole magnet.}\label{fig-quadoct}
\end{figure}
The focusing strength of the quadrupole is weaker for particles with a higher momentum, resulting in a lower oscillation frequency. This correspond to the so-called chromaticity which naturally induce a~spread in transverse frequency spread due to the momentum spread in the beam. The chromaticity can usually be controlled using sextupole magnets. In addition one often use octupole magnets, known as \textit{Landau octupoles}, which, analogously to the non-linearity of the RF wave, changes the focusing with the transverse oscillation amplitude of the particles. The fields are illustrated in Fig.~\ref{fig-quadoct}. The resulting detuning was taken as an example in previous section (Eq.~\eqref{eq-detuning-oct}).
\subsection{Accelerator design and operation}
Understanding Landau damping is key in the design and operation of accelerators. With a careful design of the components exposed to the beam aiming at minimizing their impedance and/or using active feedback systems it is possible to maximize the stability of the beam, yet it is usually not possible to suppress all unstable modes of oscillation. To maximise the performance reach of a machine several aspects should be considered to maximise Landau damping. The design of the RF system (frequency, voltage, Landau cavity) as well as its operation when longitudinal parameters evolve (e.g. due to adiabatic damping during the ramp) greatly impacts Landau damping. The optics of the machine, through the slippage factor or the chromaticity correction also impact Landau damping and therefore the performance reach of the machine. Device dedicated specifically to the enhancement and control of Landau damping are often used, such as Landau cavities or Landau octupoles. Before addressing more advanced techniques to push the performance by enhancing Landau damping in the last section, we'll discuss how various types of collective interactions may contribute to Landau damping.
\section{Non-linear collective forces}
In the first section, we derived a dispersion integral assuming a static non-linear perturbation of the~Hamiltonian (Eq.~\eqref{eq-NLPerturbation}) which describes well the impact of an external source of field such as RF cavities or magnets. However collective forces may dynamically change with the beam properties. The space-charge force for example follows the oscillation of the beam. The in-phase oscillation of the two beams in a collider (aka the beam-beam $\sigma$-mode) behaves similarly, whereas the out-of-phase oscillation (beam-beam $\pi$-mode) leads to oscillation of the non-linear field simultaneously to the beam oscillation. In all these configurations the understanding of Landau damping differs from the static case. Their complexity is such that a dispersion integral does not necessarily exists. It is quite common to rely on particle tracking simulations to evaluate the strength of Landau damping when complicated collective forces are involved.
\subsection{Space-charge}
\begin{figure}
 \begin{center}
 \includegraphics[width=0.8\linewidth]{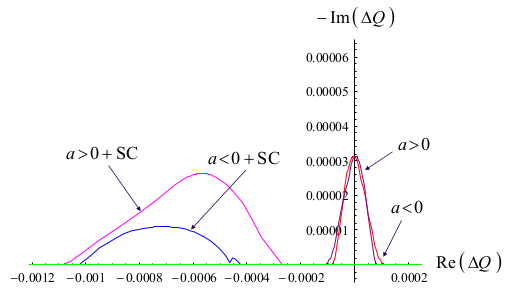}
 \end{center}
\caption{Comparison of the stability diagram caused by octupole magnets characterised by the detuning coefficient $a$ with and without space-charge (respectively the blue/pink and the red/purple line)~\cite{eliasSC}.}\label{fig-SDSpaceCharge}
\end{figure}
\begin{figure}
 \begin{center}
 \includegraphics[width=0.5\linewidth]{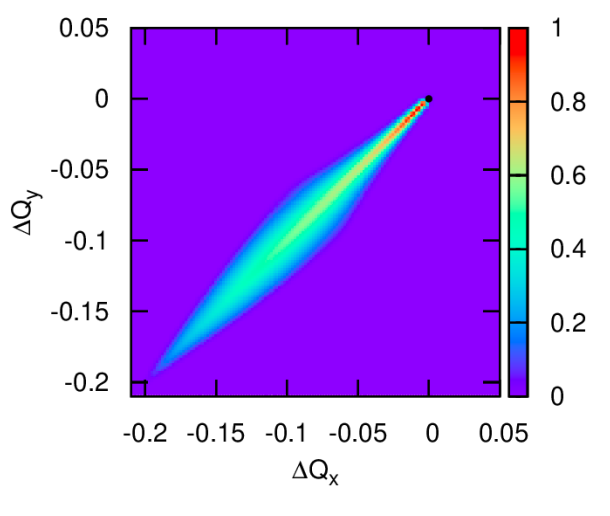}
 \end{center}
\caption{Density of the particle's frequency distribution caused by space-charge~\cite{kornilov}.}\label{fig-necktie}
\end{figure}

In the case of space-charge forces a dispersion integral was obtained in the form~\cite{eliasSC}
\begin{equation}
 \int dJ_xdJ_y \dfrac{J_x\dfrac{\partial f_0}{\partial J_x}\left(\Delta Q_{n}^x-\Delta Q^x_{SC}(J_x,J_y)\right)}{Q^x-Q^x_0-\Delta Q^{x}(J_x,J_y)-\Delta Q^x_{SC}(J_x,J_y)-nQ_s} = -1.
\end{equation}
While more complicated, the structure is comparable to the simplified example (Eq.~\eqref{eq-dispersion}). The two transverse degrees of freedom were introduced and the coherent tune shift of mode $n$ ($\Delta Q_{n}^x$) now appears inside the integral, along with the detuning caused by space-charged $\Delta Q^x_{SC}(J_x,J_y)$. The detuning was decomposed between the space-charge contribution, which appears both at the nominator and the~denominator, and the other contributions (e.g. Landau octupoles) which appear only at the denominator ($\Delta Q^{x}(J_x,J_y)$). This difference reflect that the former has a dynamic behaviour with the beam position, whereas the latter is static. The corresponding stability diagrams are fundamentally different, as shown in Fig.~\ref{fig-SDSpaceCharge}. Whereas the stability diagram without space-charge is centered around 0, the stability diagram with space-charge is shifted towards negative tune shifts. If the real tune shift of the mode 0 is small (w.r.t the space-charge tune shift), which is often the case, then the shift of the stability diagram results in a~loss of Landau damping and consequently an instability. This mechanism of loss of Landau damping can be understood in terms of wave-particle interaction. Indeed, looking at the tunes of individual particles under the influence of the space-charge force, we obtain the so-called tune footprint in Fig.~\ref{fig-necktie}. Particles with a low action may be found at the bottom left of the plot, while particles oscillating with a~higher amplitude see their frequency shift towards the unperturbed tune marked with a black point. As we assumed that the tune shift caused by the impedance is small and thus the coherent mode frequency is close to the unperturbed tune, the core of the particle distribution (low amplitude) is shifted away from it. This prevents any interaction between the coherent mode and the individual particles and thus leads to a loss of Landau damping.
\subsection{Beam-beam}
\begin{figure}
 \begin{center}
 \includegraphics[width=0.4\linewidth]{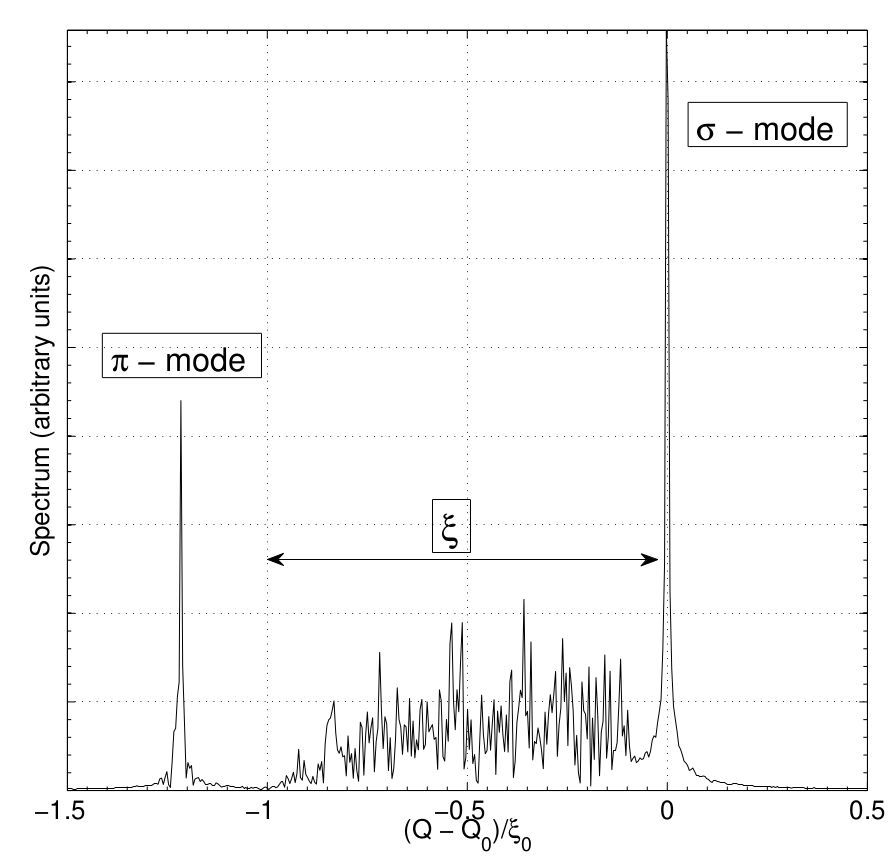}
 \end{center}
\caption{Oscillation spectrum of two bunches colliding head-on. The two main modes of oscillation correspond to in-phase or out-of-phase oscillation of the two beams (respectively the $\sigma$ and $\pi$ modes). The strength of the~beam-beam force is characterised by the so-called beam-beam parameter $\xi$~\cite{tatianaPhD}.}\label{fig-beambeam}
\end{figure}
No dispersion integrals were derived for the rigid beam-beam modes, however it could be shown that both the $\sigma$- and $\pi$-modes can lose Landau damping through a mechanism comparable to space-charge~\cite{alexahin}. This is illustrated in Fig.~\ref{fig-beambeam}, where the peaks associated with the two coherent modes of oscillation are clearly visible. In between the two modes a certain level of noise is observed in the spectrum corresponding to the oscillation frequency of the individual particles. As for space-charge, the core of the~particle distribution is shifted down in frequency (or up, for colliders featuring particles with opposite electric charge), away from the $\sigma$-mode. The $\pi$-mode is shifted even further in frequency, such that the~interaction between the mode and the individual particles is also prevented, thus suppressing Landau damping.

If the coherent beam-beam modes are suppressed (e.g. with an active feedback), the dispersion integral derived for weak head-tail modes (Eq.~\eqref{eq-dispersion}) still holds. Yet one needs to take into account that the beam-beam interactions also drive a detuning with amplitude that adds to the static ones (e.g. Landau octupoles). The different contributions may interfere favourably but also they may cancel each other and result in a loss of Landau damping~\cite{vos,PhysRevSTAB.17.111002}.
\section{Advanced techniques}
\begin{figure}
 \begin{center}
 \subfloat[Weakly non-linear motion]{
 \includegraphics[width=0.3\linewidth]{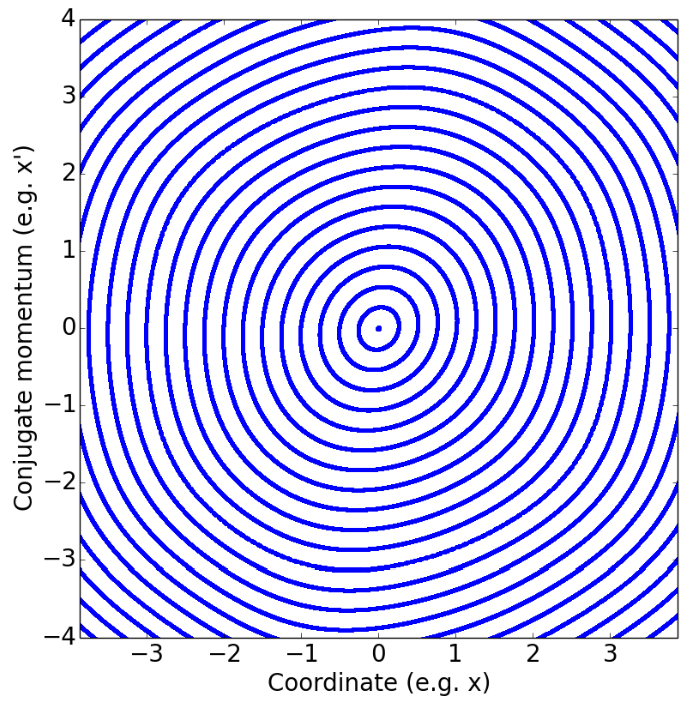}\label{fig-pointcarre-weak}}
 \qquad
  \subfloat[Strongly non-linear motion]{
 \includegraphics[width=0.3\linewidth]{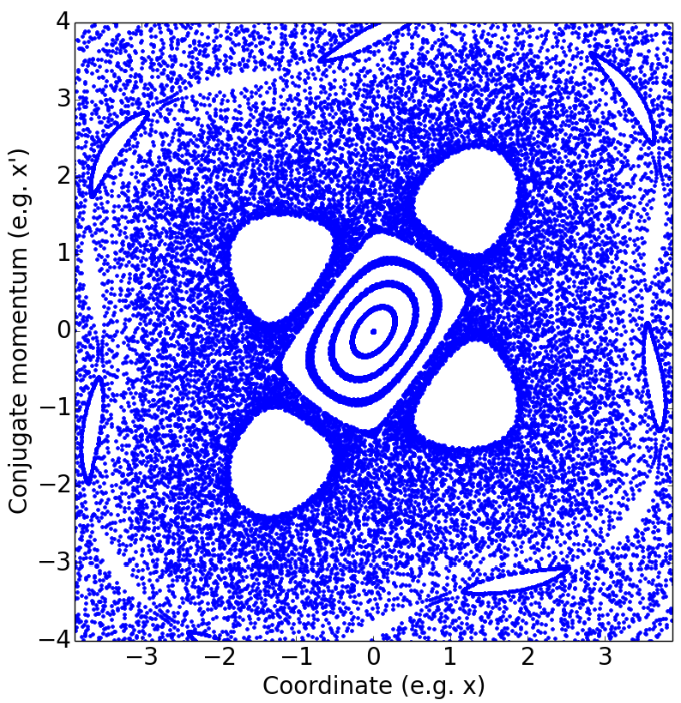}\label{fig-pointcarre-strong}}
 \end{center}
\caption{Example of particle trajectories in non-linear fields.}\label{fig-pointcarre}
\end{figure}
Whereas non-linear forces and the related frequency spread is seen as beneficial for the stability of collective motion, it is usually detrimental for long term stability of the trajectory of individual particles. The trajectories of a set of particles in phase space in a weakly and strongly non-linear configuration are shown in Fig.~\ref{fig-pointcarre}. In the weakly non-linear case, there exists a frequency spread and the trajectories are slightly distorted, thus yielding a good balance between stability of collective modes and stability of individual particles. This balance translate into a stable beam with a long lifetime. In contrary in the~second case the trajectories are heavily distorted leading to chaos and thus strong diffusion of the~particles towards higher oscillation amplitude. Eventually those particles will hit the physical aperture resulting in beam losses and consequently a low beam lifetime. Thus we understand that Landau damping is usually available in a limited amount. In addition it is not only limited by technical limitations, such as the~strength of Landau cavities or Landau octupoles, but also by fundamental mechanisms of particle losses. Several approaches are studied nowadays to maximize Landau damping while keeping the~mechanisms of diffusion under control by introducing other types of non-linear forces.
\subsection{Electron lens}
An electron lens is a device that generates an electron beam which propagates against the main beam in a~short section of the accelerator. The force experienced by the particles in the main beam can be shaped by adjusting the distribution of the electron beam. A simple Gaussian beam distribution is already sufficient to increase significantly the tune spread, with respect to octupole magnets, while maintaining stable trajectories~\cite{PhysRevLett.119.134802}. To achieve this improvement, the key feature of the Gaussian electron lens over non-linear magnets is the fact that they induce a large tune spread for particles with a small oscillation amplitude, while the amplitude detuning vanishes at high oscillation amplitude. Since the beam distribution is more populated at low oscillation amplitude, they provide stronger Landau damping for a given tune spread.
\subsection{Non-linear integrable optics}
\begin{figure}
 \begin{center}
 \subfloat[Particle trajectories~\cite{PhysRevSTAB.13.084002}]{
 \includegraphics[width=0.4\linewidth]{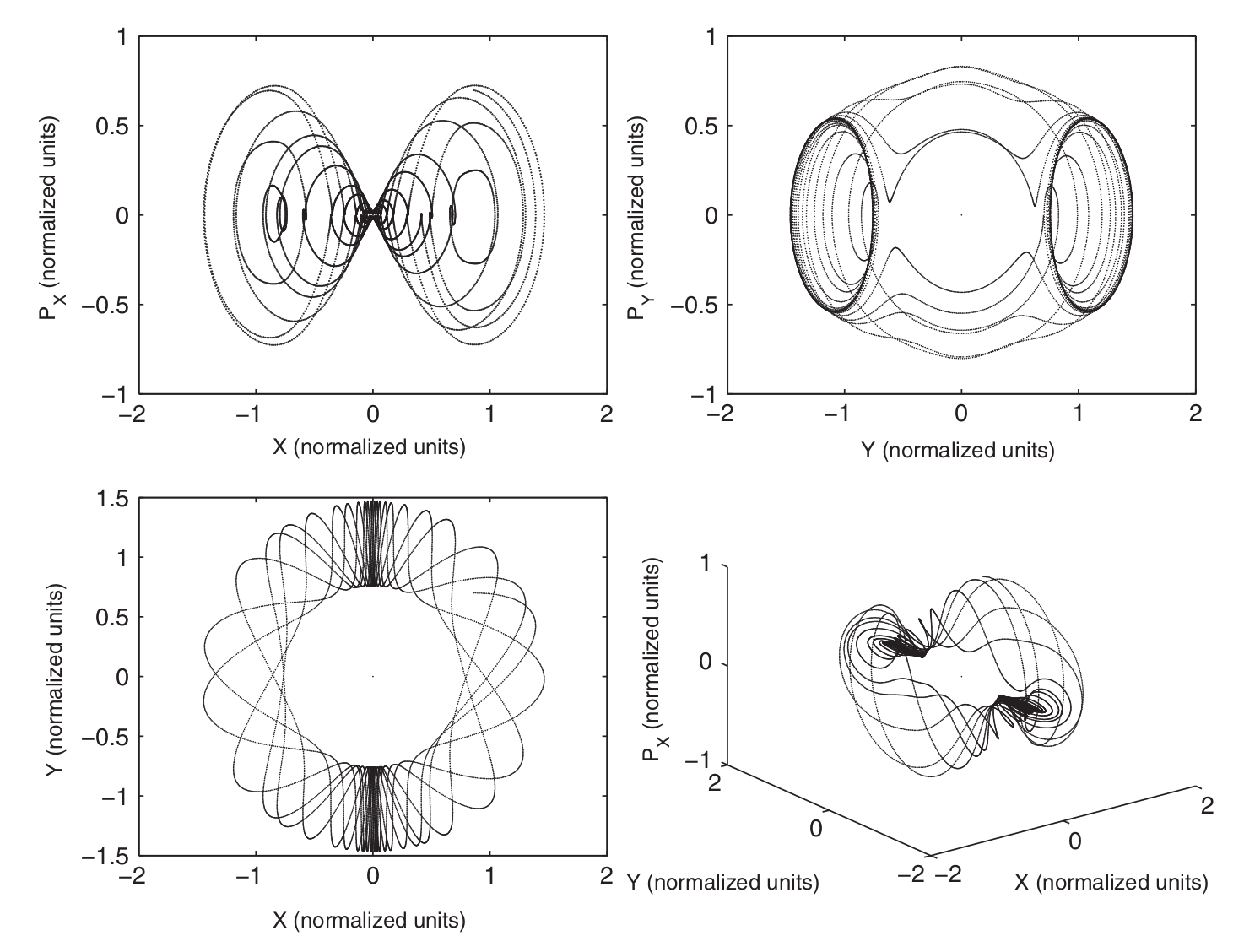}\label{fig-NLIOtrajectories}}
 \qquad
  \subfloat[Individually powered octupoles at IOTA~\cite{IOTA}]{
 \includegraphics[width=0.45\linewidth]{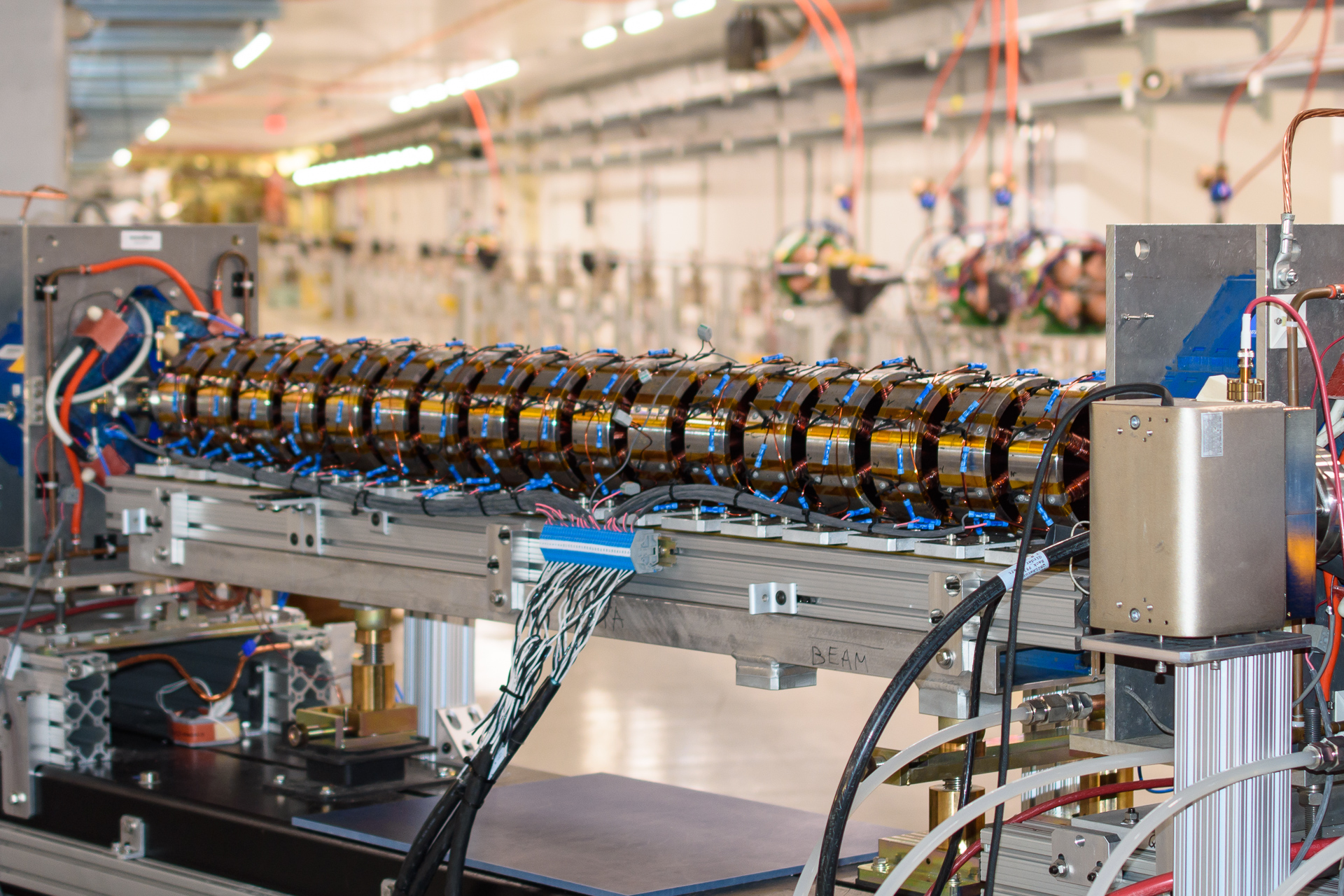}\label{fig-IOTAOct}}
 \end{center}
\caption{Examples of particle trajectories in non-linear integrable optics and a system to achieve it.}\label{fig-NLIO}
\end{figure}
Dynamical systems are called integrable when they feature conserved quantities. If the motion of the~particles in an accelerator is integrable, the existence of conserved quantities (such as the action) implies that the trajectories are stable on long time scale. While it is clear that perfectly linear accelerator features conserved quantities (e.g. the transverse and longitudinal actions $J_x$, $J_y$ and $J_z$), those quantities are usually not exactly conserved in machines featuring non-linearities. It is possible to generate non-linear forces such that there still exists preserved quantities, we then obtain so-called non-linear integrable optics~\cite{PhysRevSTAB.13.084002}. Examples of stable trajectories in a non-linear integrable optics are shown in Fig.~\ref{fig-NLIO}. Such optics can be achieved by introducing special devices in an otherwise standard accelerator lattice, for example by choosing a specific density profile of an electron lens~\cite{Nagaitsev_2021} or power a series of octupoles in a specific way~(Fig.~\ref{fig-IOTAOct}).
\subsection{Radio-frequency quadrupole}
\begin{figure}
 \begin{center}
 \subfloat[Field map~\cite{PhysRevAccelBeams.20.082001}]{
 \includegraphics[width=0.4\linewidth]{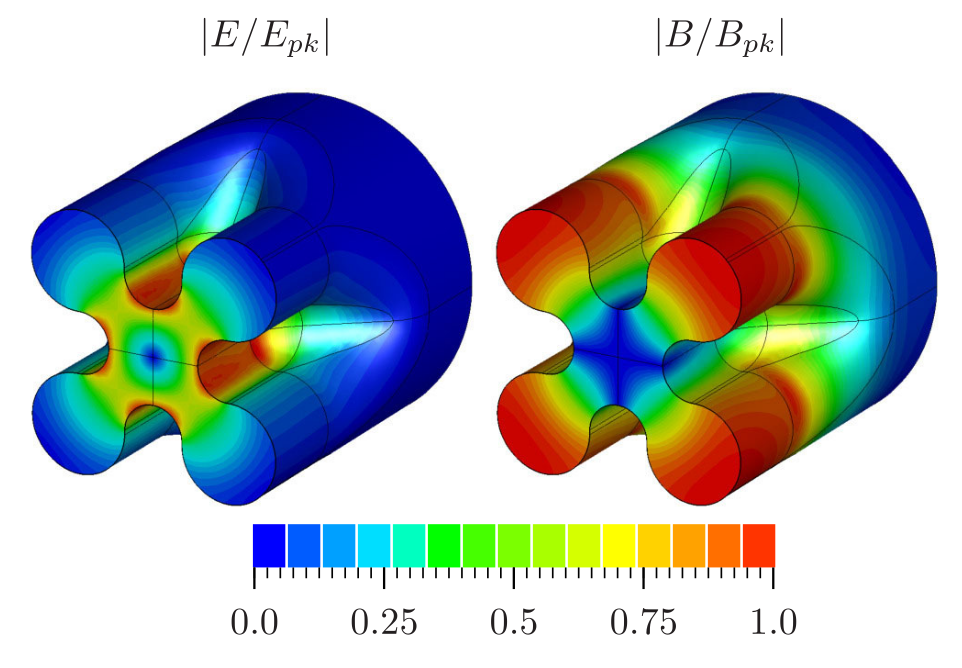}\label{fig-RFQdevice}}
 \qquad
  \subfloat[Stability diagram~\cite{schenk}]{
 \includegraphics[width=0.4\linewidth]{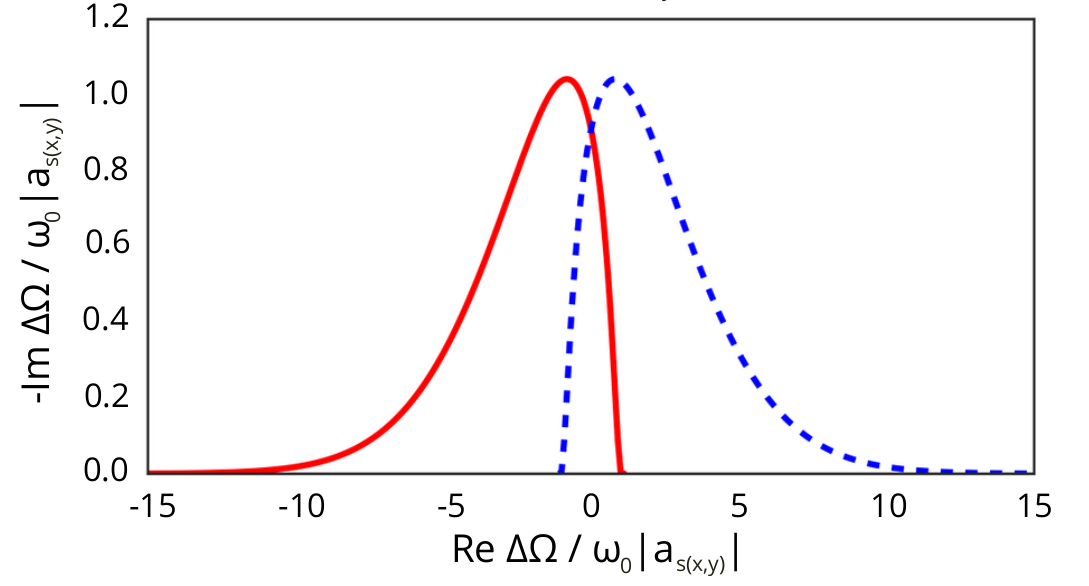}\label{fig-RFQSD}}
 \end{center}
\caption{Field map of a RF-quadrupole and the corresponding stability diagram.}\label{fig-RFQ}
\end{figure}
A radio-frequency quadrupole feature a transverse quadrupole field that varies fast in time (Fig.~\ref{fig-RFQdevice}), such that the focusing strength varies along the bunch. This device therefore introduces a transverse tune spread depending on the longitudinal positions of the particles. In that case the dispersion integral takes yet a different form~\cite{PhysRevSTAB.17.011001} resulting a stability diagram shown in Fig.~\ref{fig-RFQSD}. While the motion is not integrable, as for Landau octupole, the transverse Landau damping that can be obtained without compromising the stability of single particle trajectories may still be improved~\cite{schenk}. 

\subsection{Acknowledgment}
I would like to thank W. Herr for valuable comments on this manuscript.

\bibliographystyle{apsrev4-1-title}
\bibliography{CAS_LandauDamping}

\end{document}